\documentclass[conference]{IEEEtran}\usepackage{cite}
\usepackage{amsmath,amssymb,amsfonts}
\usepackage{graphicx}
\usepackage{textcomp}
\usepackage{url}
\usepackage{xcolor}
\usepackage{tikz}

\newcommand\copyrighttext{%
  \footnotesize \textcopyright \the\year{} IEEE. Personal use of this material is permitted. Permission from IEEE must be obtained for all other uses, including reprinting/republishing this material for advertising or promotional purposes, collecting new collected works for resale or redistribution to servers or lists, or reuse of any copyrighted component of this work in other works.}

\newcommand\copyrightnotice{%
\begin{tikzpicture}[remember picture,overlay]
\node[anchor=south,yshift=10pt] at (current page.south) {\fbox{\parbox{\dimexpr0.75\textwidth-\fboxsep-\fboxrule\relax}{\copyrighttext}}};
\end{tikzpicture}%
}

\def\BibTeX{{\rm B\kern-.05em{\sc i\kern-.025em b}\kern-.08em
    T\kern-.1667em\lower.7ex\hbox{E}\kern-.125emX}}

\begin{document}

\title{The Kubernetes Network Driver Model: A Composable Architecture for High-Performance Networking}

\author{\IEEEauthorblockN{Antonio Ojea}
\IEEEauthorblockA{\textit{Google Kubernetes Engine} \\
\textit{Google}\\
aojea@google.com}
}

\maketitle

\copyrightnotice

\begin{abstract}
Traditional Kubernetes networking struggles to meet the escalating demands of AI/ML and evolving Telco infrastructure. This paper introduces Kubernetes Network Drivers (KNDs), a transformative, modular, and declarative architecture designed to overcome current imperative provisioning and API limitations. KNDs integrate network resource management into Kubernetes' core by utilizing Dynamic Resource Allocation (DRA), Node Resource Interface (NRI) improvements, and upcoming OCI Runtime Specification changes. Our DraNet implementation demonstrates declarative attachment of network interfaces, including Remote Direct Memory Access (RDMA) devices, significantly boosting high-performance AI/ML workloads. This capability enables sophisticated cloud-native applications and lays crucial groundwork for future Telco solutions, fostering a "galaxy" of specialized KNDs for enhanced application delivery and reduced operational complexity.
\end{abstract}

\section{Introduction}

The evolution of Kubernetes has been driven by the need to manage increasingly complex and diverse workloads. While the initial focus was on stateless applications, the platform is now a cornerstone for stateful, performance-sensitive tasks, including large-scale AI/ML training and inference. These workloads demand direct, low-latency access to specialized hardware, such as GPUs and RDMA-capable network interfaces. However, the mechanisms originally designed for networking and management of hardware resources, such as the Container Networking Interface specification (CNI) \cite{CNISpec} and the Kubernetes extended resources (Device Plugins) \cite{k8s_device_plugin}, were not designed for the intricate requirements of this new class of hardware, leading to significant challenges in expressiveness, composability, and operational simplicity.

\begin{figure}
    \centering
    \includegraphics[width=0.75\linewidth]{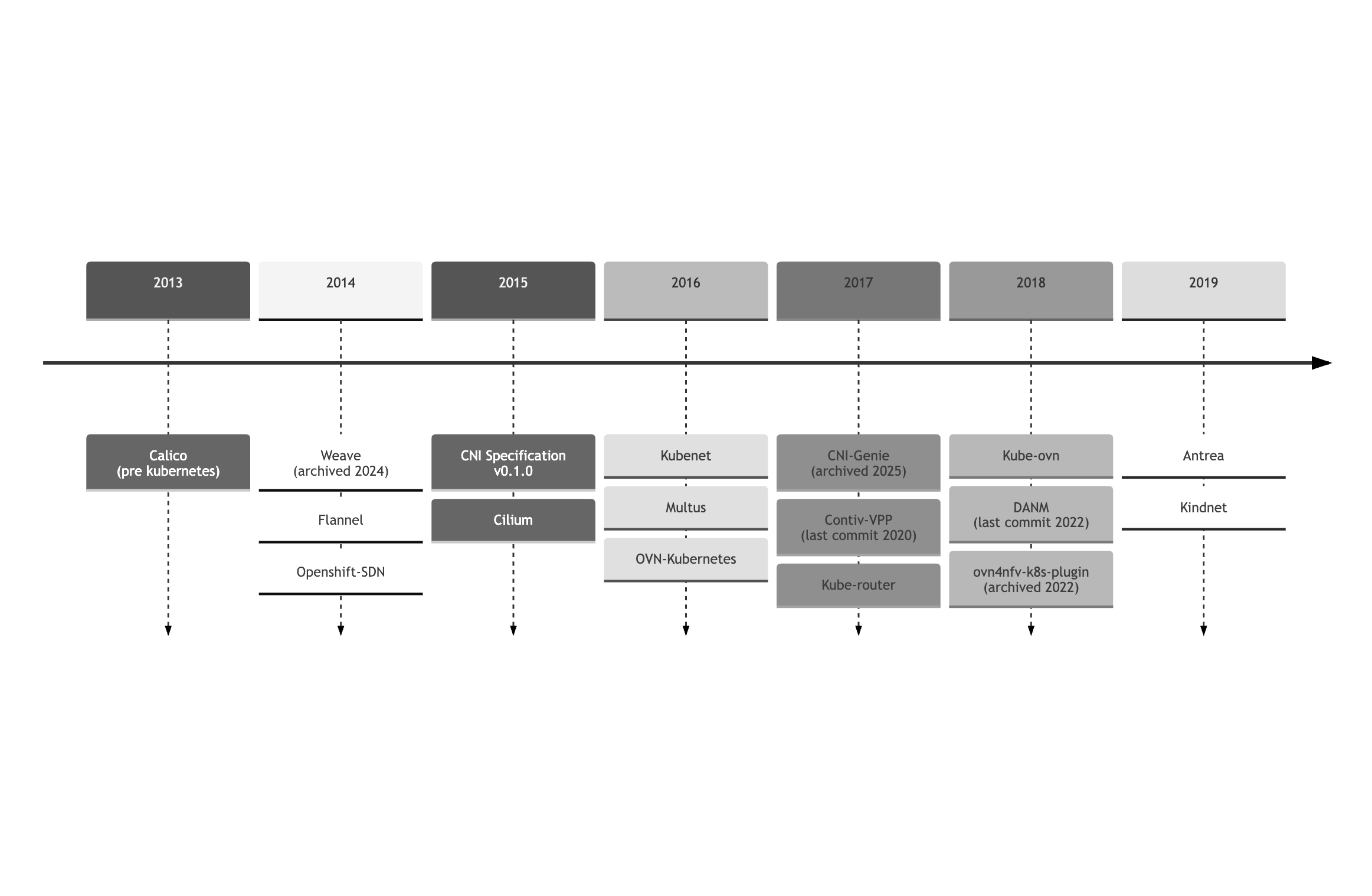}
    \caption{Timeline of Key Kubernetes Networking Projects.}
    \label{fig:oss-projects}
\end{figure}

The history of Kubernetes networking, illustrated in Figure \ref{fig:oss-projects}, shows a rapid proliferation of projects attempting to solve various networking challenges.

Notably, many of the projects specifically targeting advanced networking use cases for telecommunications (e.g., CNI-Genie, DANM, ovn4nfv-k8s-plugin) have since been abandoned or archived. This has left the ecosystem dominated by two primary architectural patterns: large, monolithic "thick plugins" like Calico \cite{calico} or Cilium \cite{cilium} that implement the entirety of network functionality, and meta-plugins like Multus \cite{multus}, which provide a multiplexing layer to attach multiple, simple CNI interfaces. Neither of these approaches adequately addresses the need for expressive, topology-aware management of high-performance hardware.

This paper identifies the architectural limitations of these legacy models and proposes a new paradigm: the Kubernetes Network Driver (KND) Model. The KND model shifts away from these solutions towards a system of independent, composable drivers that leverage first-class Kubernetes APIs. We demonstrate that by using Dynamic Resource Allocation (DRA) \cite{KEP4381} for expressive resource claims and the Node Resource Interface (NRI) \cite{ContainerdNRI} for runtime composability, we can build a more robust, efficient, and simpler system for high-performance networking.

To validate this model, we introduce DraNet \cite{GoogleDranet}, a reference implementation of the KND model. It is designed to manage host network interfaces as dynamic, first-class resources within Kubernetes, including specialized hardware like RDMA devices.

\section{Background and Motivation}

In the early days of Kubernetes, the Container Network Interface (CNI) was chosen over Docker's Container Network Model (CNM) for its simplicity and flexibility \cite{hockin2016}. CNI provided a focused interface for network plugins, allowing a diverse ecosystem of networking solutions to flourish. However, as user demands grew for features like network policies, service discovery and multiple interfaces, the simple CNI specification led to the rise of "thick plugins." 

The Kubernetes community recognized the shortcomings of the CNI model for advanced use cases. Discussions within SIG-Network led to formal proposals to address multi-networking natively, most notably through KEP-3698 "Multi-Network for Pods"\cite{KEP3698} and the subsequent KEP-4410 "KNI: Kubernetes Network Interface" \cite{KEP4410}. These enhancement proposals accurately identified the core problem and explored potential solutions. However, neither initiative progressed to a formal implementation, leaving a significant architectural gap in the platform.

\begin{figure}
    \centering
    \includegraphics[width=1\linewidth]{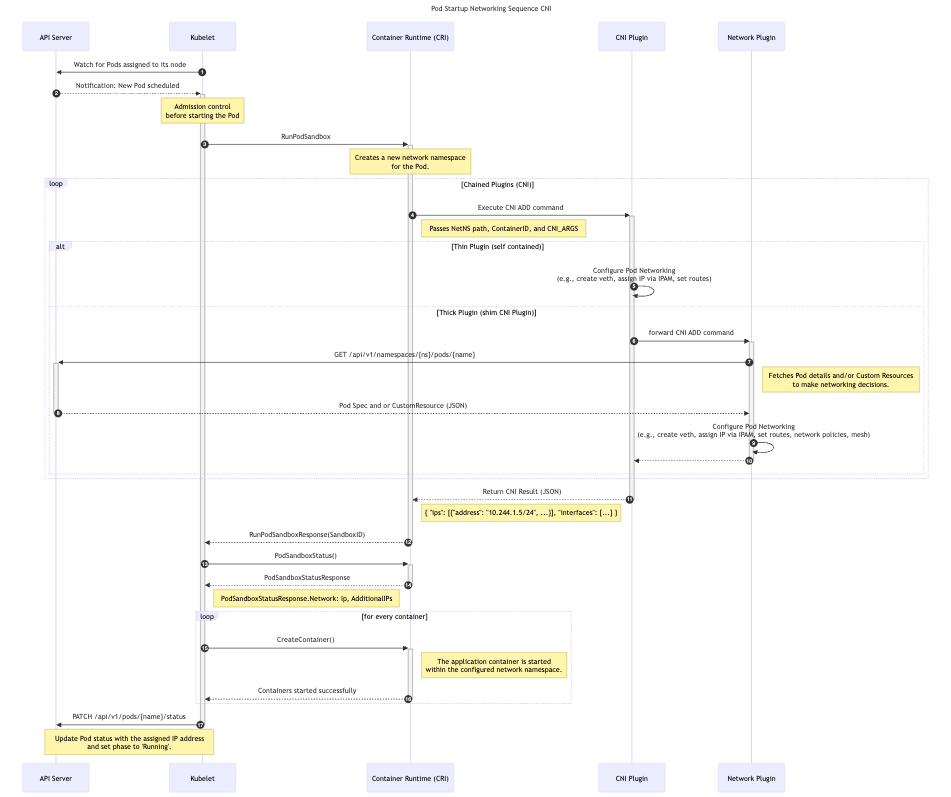}
    \caption{Pod Startup Networking Sequence with CNI}
    \label{fig:cni-startup}
\end{figure}

This lack of a native solution compelled the community to adopt workarounds extended Kubernetes networking via Custom Resource Definitions (CRDs) and complex node agents, often using a shim CNI binary that delegated work to a long-running daemon. This pattern, while functional, introduced latency and a dependency requiring Kubernetes API server lookups during the critical pod startup path, as shown in Figure \ref{fig:cni-startup}. It also creates a critical lifecycle mismatch between the CNI binary on disk and its corresponding daemon. The container runtime can execute the binary at any time, but if the daemon process is restarting or has crashed, the operation will fail after a lengthy timeout due to this lack of coordination.

\begin{figure}
    \centering
    \includegraphics[width=1\linewidth]{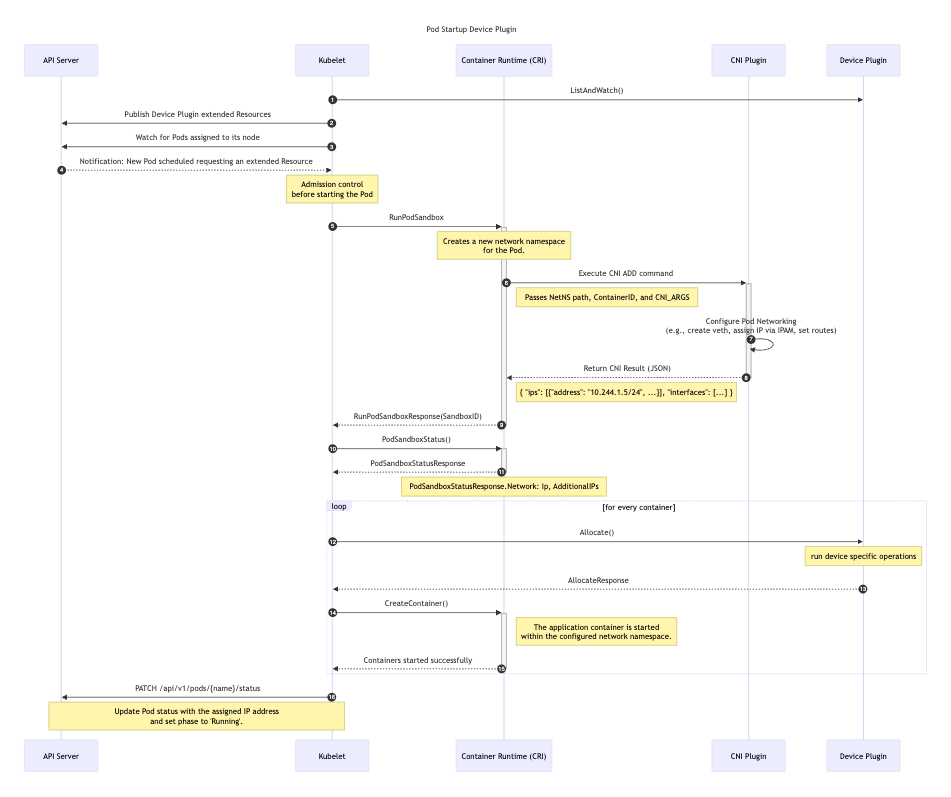}
    \caption{Pod Startup Networking Sequence with CNI and Device Plugin}
    \label{fig:cni-dev-plugin-startup}
\end{figure}

The problem was amplified when exposing specialized hardware like RDMA NICs. The standard solution involved a fragile composition of disparate components with a more complex startup sequence as shown in Figure \ref{fig:cni-dev-plugin-startup}: a CNI meta-plugin like Multus, a hardware-specific Device Plugin and a dedicated CNI plugin to configure the device. This approach suffers from several fundamental flaws.

First, as CNI is a property delegated to the container runtime \cite{K8sContainerRuntimes}, it fundamentally lacks the expressiveness needed to influence the Kubernetes scheduler. This architectural separation makes it impossible for the control plane to make intelligent placement decisions based on network topology or capabilities. This limitation is typically compensated for through out-of-band mechanisms, such as daemons that discover network features and label node objects, or by using the Device Plugin framework itself as a proxy for network resources.

However, this leads to the second flaw: the Device Plugin framework is purely quantitative, advertising a count of resources, and is incapable of expressing the rich qualitative attributes or topological relationships (like PCI locality) essential for performance. Third, there is a severe scope mismatch, as device plugins operate on a per-container basis, while network interfaces are a pod-level resource. Finally, there is no native synchronization between the device plugin's allocation and the CNI plugin's configuration, leading to complex and brittle implementations that rely on passing state through annotations.

\section{The Kubernetes Network Driver (KND) Model}

To address these limitations, we propose the Kubernetes Network Driver (KND) model, built upon two modern, first-class Kubernetes and Controller Runtime APIs: Dynamic Resource Allocation (DRA) and the Node Resource Interface (NRI).

\begin{figure}
    \centering
    \includegraphics[width=1\linewidth]{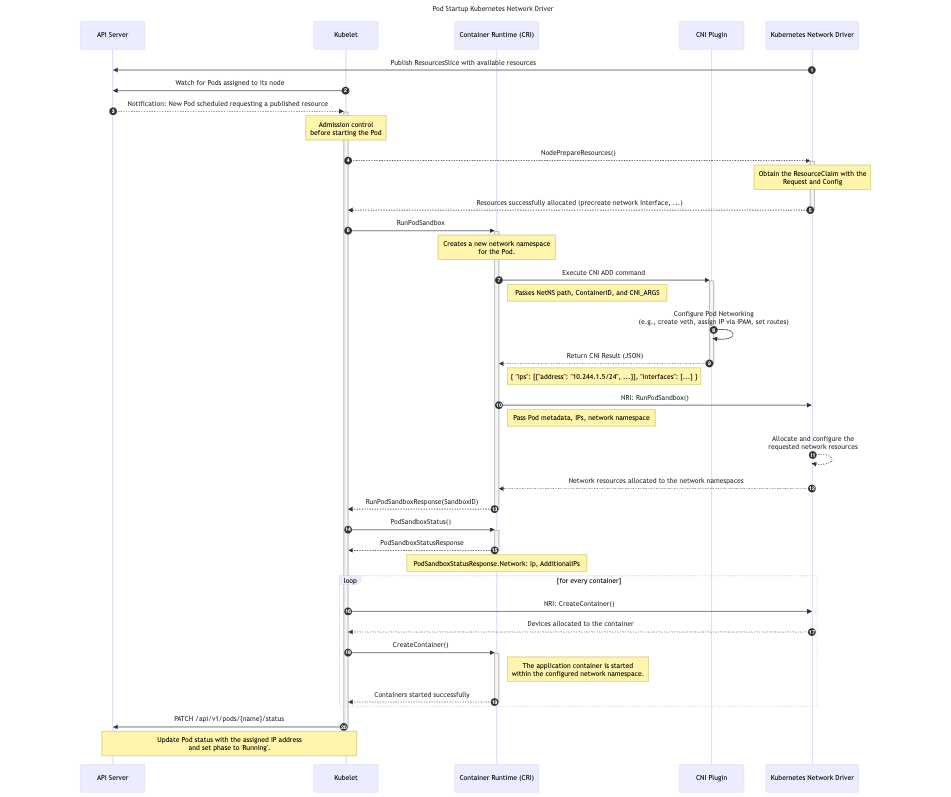}
    \caption{Pod Startup Networking Sequence with KND}
    \label{fig:knd-startup}
\end{figure}

\subsection{Dynamic Resource Allocation (DRA): Expressive, Topology-Aware Scheduling}
DRA is the designated successor to the device plugin framework, designed to manage complex hardware resources natively within Kubernetes . It overcomes the limitations of its predecessor by introducing several key features:
\begin{itemize}
    \item \textbf{Richer Resource Profiles:} Drivers can expose any resource, be it physical, virtual, or purely logical, with both quantitative and qualitative attributes. This allows a driver, for example, to publish not just the existence of a physical NIC, but also its NUMA node and PCI root address. Crucially, the same mechanism can be used to model more abstract resources, such as an SR-IOV Virtual Function or even a provisioned network service like an MPLS tunnel, making them discoverable and schedulable by Kubernetes.
    \item \textbf{Expressive User Intent:} Users request resources via ResourceClaim objects, using the powerful Common Expression Language (CEL) \cite{cel_spec} for selection. This enables topology-aware scheduling, where a user can request a GPU and a NIC that share the same PCI root, allowing the Kubernetes scheduler to find a node that satisfies this constraint, thereby eliminating a major performance bottleneck.
    \item \textbf{Decoupled Lifecycle and Embedded Parameters:} DRA introduces a NodePrepareResources hook, allowing a driver to perform slow setup operations before the pod's critical startup phase. Crucially, the ResourceClaim object can carry opaque configuration parameters directly to the driver during this hook. This mechanism enables a "push" model, where all necessary information is provided upfront, completely removing the need for the driver to connect back to the API server and thereby eliminating a major source of latency and potential race conditions, shown in Figure \ref{fig:knd-startup}.
\end{itemize}

\subsection{Node Resource Interface (NRI): Composable Runtime Hooks}
NRI provides a generic, event-driven plugin architecture that allows multiple independent drivers to hook into the container runtime lifecycle. This solves the composability problem of CNI chaining, where multiple plugins are executed in a rigid, sequential pipeline. With NRI, different drivers (e.g., a GPU driver and a network driver) can subscribe to pod lifecycle events (like RunPodSandbox or CreateContainer) and act in parallel and without direct dependencies. This enables a clean separation of concerns, where a driver can handle network attachment at the pod level while the other driver handles GPU setup at the container level. Crucially, these hooks are not just triggers; they are context-aware, providing the driver with all the necessary information to perform its operations. For instance, recent enhancements to the NRI specification ensure that plugins receive the pod's full network state, including its assigned IP addresses \cite{ContainerdNRI119}.

\subsection{OCI Spec Simplification}
Further simplification is achieved by leveraging recent additions to the Open Container Initiative (OCI) \cite{OCIRuntimeSpec} runtime specification that allow for the declarative attachment of network interfaces \cite{OCIRuntimeSpec1271}. This allows network drivers to simply instruct the container runtime to move a prepared interface into the pod's namespace, offloading the privileged, low-level netlink operations to the runtime itself and reducing the capabilities required by the driver.

\begin{figure}
    \centering
    \includegraphics[width=0.75\linewidth]{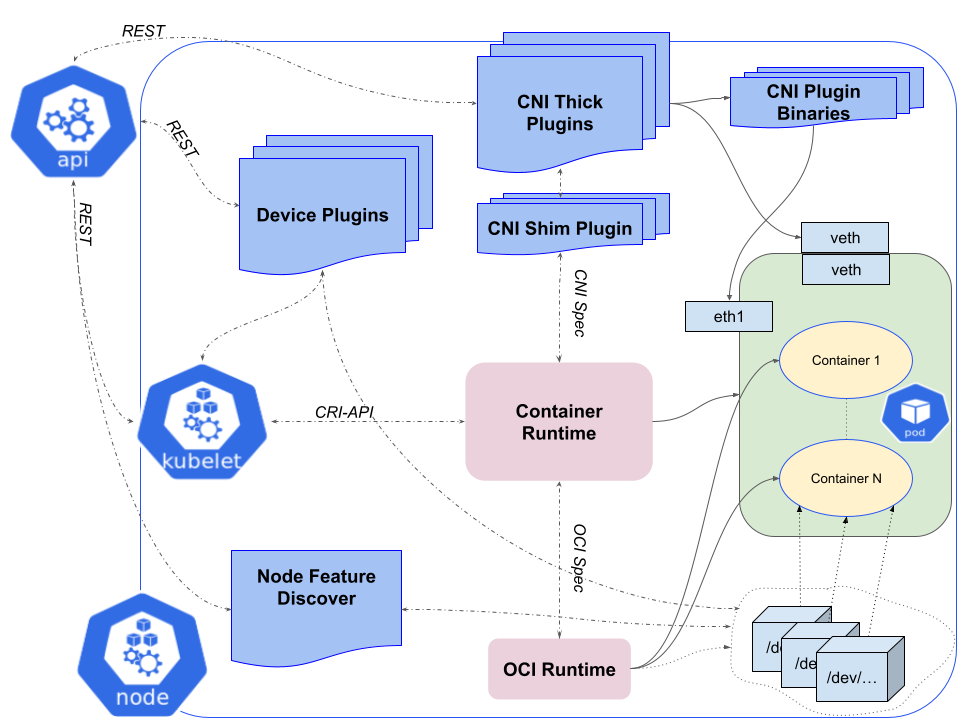}
    \caption{Classic CNI and Device Plugin Architecture}
    \label{fig:cni-dev-arch}
\end{figure}

\begin{figure}
    \centering
    \includegraphics[width=0.75\linewidth]{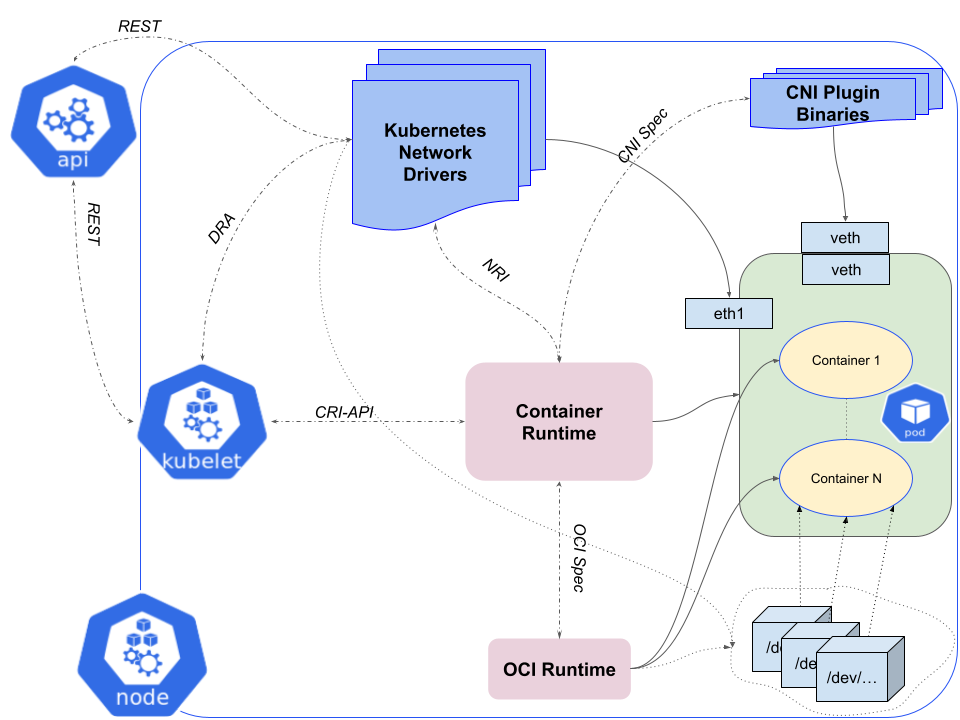}
    \caption{Kubernetes Network Driver Architecture}
    \label{fig:knd-arch}
\end{figure}

\section{DraNet: A Reference Implementation}

To provide a concrete example of the KND model, we developed DraNet, an Open Source reference implementation designed to manage the lifecycle of entire network interfaces as first-class Kubernetes resources. This framework is capable of provisioning any raw network device, from standard host interfaces to specialized high-performance RDMA hardware, through the same native Kubernetes APIs.

\subsection{Architectural Simplicity and Operational Overhead}

A key advantage of the KND model demonstrated by DraNet is the drastic reduction in operational complexity. The traditional model (Figure \ref{fig:cni-dev-arch}) for RDMA required a fragile, three-component chain (e.g., Multus + SR-IOV Device Plugin + RDMA CNI).

By contrast, the KND model enables a system of independent, composable drivers (Figure \ref{fig:knd-arch}). When used for GPU-aligned networking, the setup involves only two components:
\begin{enumerate}
    \item The NVIDIA DRA GPU Driver\cite{nvidia_dra_driver}: Manages the GPU lifecycle.
    \item The DraNet Driver: Manages the network resource lifecycle.
\end{enumerate}

Both drivers interact with the same standard Kubernetes API in parallel, without direct dependencies. This architectural purity simplifies installation, reduces potential points of failure, and makes the system easier to debug and maintain.

\subsection{End-to-End Workflow}

\begin{figure}
    \centering
    \includegraphics[width=1\linewidth]{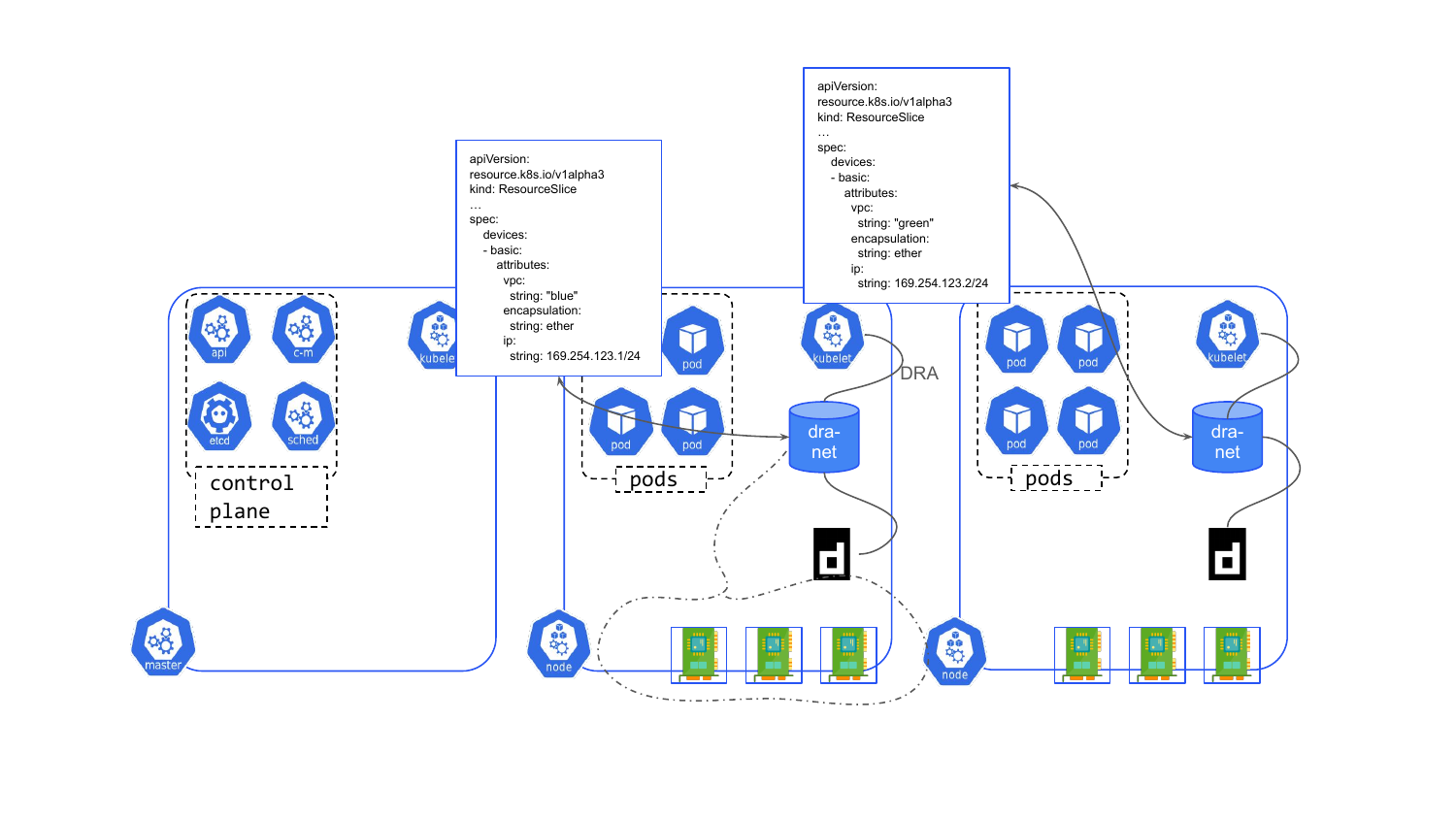}
    \caption{DraNet Deployment}
    \label{fig:dranet-arch}
\end{figure}

The DraNet workflow exemplifies the KND model's elegance (Figure \ref{fig:dranet-arch}):
\begin{enumerate}
    \item \textbf{Discovery:} The DraNet daemon on each node discovers network interfaces and their topological attributes (PCI root, NUMA node) and publishes them as ResourceSlices API objects.
    \item \textbf{Claiming \& Scheduling:} A user creates a ResourceClaim API object requesting an RDMA NIC. The Kubernetes scheduler, using the rich attributes provided by DraNet, finds an optimal node for placement.
    \item \textbf{Preparation (DRA Hook):} The kubelet calls NodePrepareResources. DraNet prepares the specific device and caches the user's configuration.
    \item \textbf{Attachment \& Configuration (NRI Hooks):} During RunPodSandbox, DraNet moves the network interface into the pod's namespace. During the subsequent CreateContainer hook, it presents the required RDMA character devices (e.g., /dev/infiniband/uverbsN) to the container.
\end{enumerate}

This clean separation ensures pod-level network resources are handled distinctly from container-level device access, fully realizing the KND model's design goals.

\section{Performance Evaluation}

To validate the performance benefits of the topology-aware scheduling enabled by the KND model, we conducted a series of benchmarks on Google Cloud. The experiments were designed to measure two key aspects: the operational efficiency of the model via pod startup latency, and the application-level network performance under both optimal and suboptimal hardware configurations.

\subsection{Experiment Setup}
\begin{itemize}
    \item \textbf{Hardware and Software:} The testbed consisted of two Google Cloud nodes of type a4-highgpu-8g. Each node is equipped with eight NVIDIA B200 GPUs and eight Mellanox RoCE NICs. The nodes run Google Kubernetes Engine (GKE) version v1.33.1-gke.1744000 with the DynamicResourceAllocation feature gate enabled \cite{dranet_nvidia_guide}.
    \item \textbf{Workloads and Experimental Conditions:} The core of our experiment involved comparing two distinct workload configurations, each repeated 100 times to ensure statistical significance.
    \begin{enumerate}
        \item \textbf{Topologically Aligned:} This configuration used one ResourceClaimTemplate for the GPU and another for the RDMA NIC. The ResourceClaimTemplate uses a CEL selector to request an RDMA NIC that is known to be on the same PCI root as the requested GPU (e.g., gpu0rdma0 for GPU 0).
        \item \textbf{Topologically Unaligned (High Variance):} This configuration used the extended resources via the device plugin for the GPU request. It explicitly requests a specific RDMA NIC via ResourceClaim in order to guarantee the communications between both nodes. However, the GPU is requested via the traditional, non-DRA-aware device plugin mechanism. Since the device plugin has no context of the network resource claim, it randomly assigns one of the eight available GPUs on the node. This creates high performance variance, as there is only a 1-in-8 chance that the allocated GPU will be topologically aligned with the requested NIC.
    \end{enumerate}
    \item \textbf{Methodology:} For both configurations, a script deployed the StatefulSet with a Headless Service, waited for pod readiness, executed the benchmarks, and then deleted the StatefulSet. The NCCL tests \cite{nccl_tests} (all\_gather and all\_reduce) were run with robust parameters (-b 8 -e 8G -f 2 -n 100 -w 50). Each test process was configured to use only a single GPU and its corresponding single network interface (-g 1). This design choice is critical to isolate the performance of the inter-node RDMA network fabric. If multiple GPUs on the same node were used, NCCL would automatically prefer the much faster intra-node NVLink interconnect for communication, which would mask or "contaminate" the network performance results this study aims to measure.

\end{itemize}

\subsection{Pod Startup Latency}
The pod startup latency was analyzed by parsing the timestamps from the 100 pod creation events for the aligned workloads. The results demonstrate the high operational efficiency of the KND model.

\begin{table}[h!]
    \centering
    \caption{Pod Startup Latency Percentiles}
    \begin{tabular}{@{}lc@{}}
    \textbf{Percentile} & \textbf{Startup Latency (s)} \\
    P50 (Median) & 1.8 \\
    P90 & 2.1 \\
    P99 & 2.3 \\
    \end{tabular}
    \label{tab:latency}
\end{table}

\subsection{Network Performance Results}

\begin{figure}
    \centering
    \includegraphics[width=0.75\linewidth]{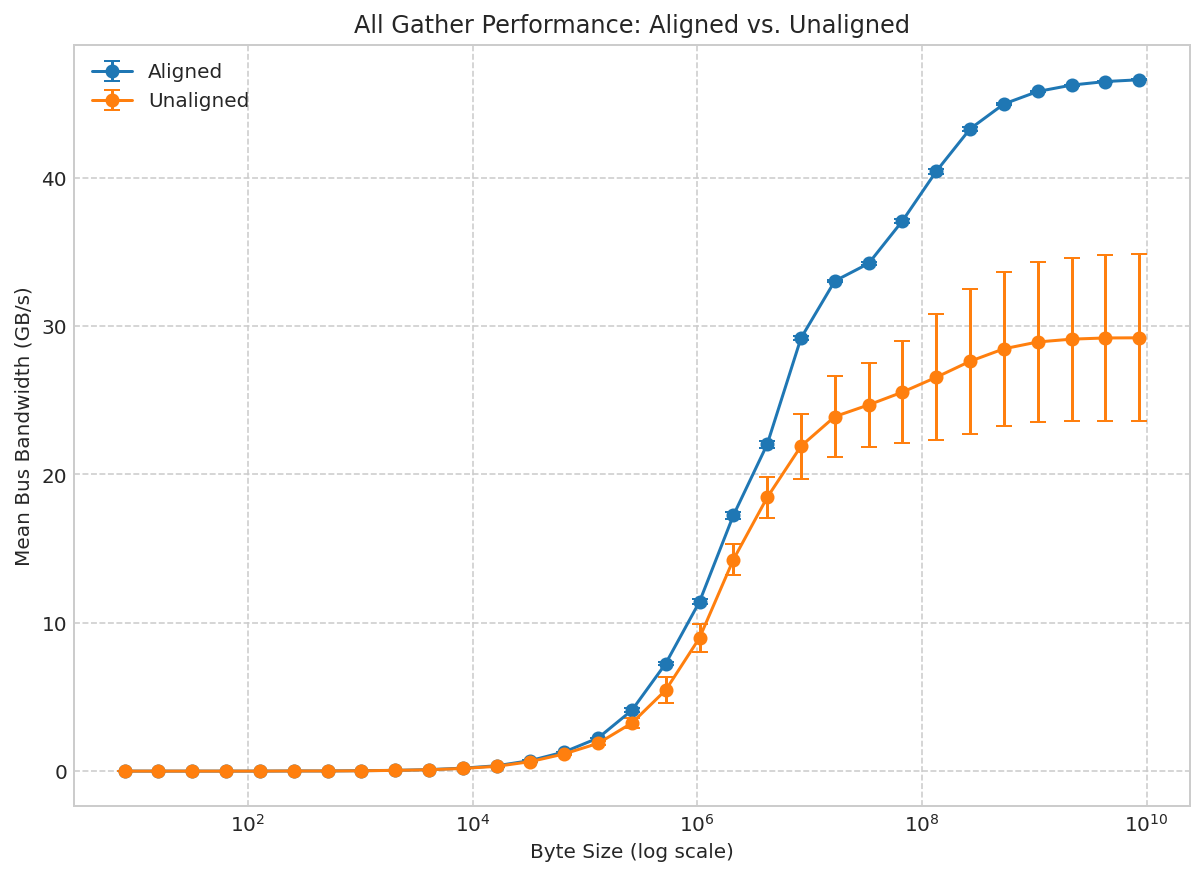}
    \caption{NCCL All Gather Benchmark}
    \label{fig:nccl_all_gather}
\end{figure}

\begin{figure}
    \centering
    \includegraphics[width=0.75\linewidth]{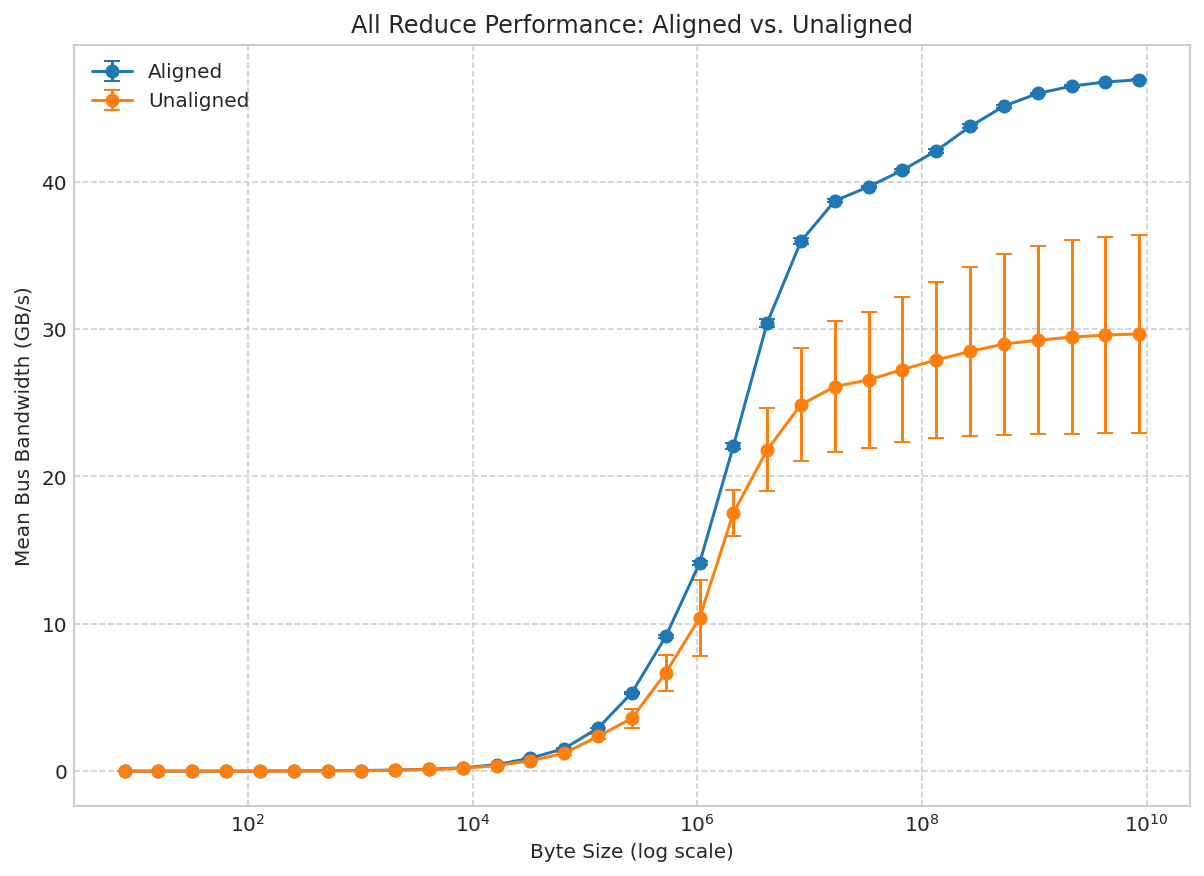}
    \caption{NCCL All Reduce Benchmark}
    \label{fig:nccl_all_reduce}
\end{figure}

The comparison between the aligned and unaligned configurations reveals a dramatic performance difference (Figures \ref{fig:nccl_all_gather} and \ref{fig:nccl_all_reduce}). Including the standard deviation (StdDev) in our results highlights the performance instability caused by the random GPU assignment in the unaligned case, which is a critical finding of this work.

\begin{table}[h!]
    \centering
    \caption{NCCL All-Gather Bus Bandwidth Summary (GB/s)}
    \begin{tabular}{lccc}
    \hline
    \textbf{Message Size} & \textbf{Aligned (Mean ± StdDev)} & \textbf{Unaligned (Mean ± StdDev)} \\
    \hline
    64 KB & 1.29 (± 0.02) & 1.16 (± 0.06) \\
    1 MB & 11.42 (± 0.19) & 8.98 (± 0.95) \\
    8 GB & 46.59 (± 0.03) & 29.20 (± 5.62) \\
    \end{tabular}
    \label{tab:all_gather_summary}
\end{table}

\begin{table}[h!]
    \centering
    \caption{NCCL All-Reduce Bus Bandwidth Summary (GB/s)}
    \begin{tabular}{lccc}
    \hline
    \textbf{Message Size} & \textbf{Aligned (Mean ± StdDev)} & \textbf{Unaligned (Mean ± StdDev)} \\
    \hline
    64 KB & 1.53 (± 0.03) & 1.21 (± 0.11) \\
    1 MB & 14.11 (± 0.13) & 10.39 (± 2.60) \\
    8 GB & 46.93 (± 0.04) & 29.68 (± 6.74) \\
    \end{tabular}
    \label{tab:all_reduce_summary}
\end{table}

The results are conclusive. The aligned configuration not only achieves significantly higher mean throughput, but also exhibits dramatically lower variance. Conversely, the unaligned configuration suffers from both lower average performance and high variance, reflecting the "lottery" of whether the randomly assigned GPU happened to be on the correct PCI root. This instability makes performance unpredictable and is unacceptable for production HPC and AI workloads.

\section{Discussion}
Our experiments clearly demonstrate that topological alignment between GPUs and NICs, enabled by the KND model, yields a significant performance increase, boosting bus bandwidth by up to 59.6\% in our all\_gather tests and a 58.1\% in our all\_reduce tests. These results confirm our hypothesis that the lack of topological awareness in traditional device management models is a primary source of performance bottlenecks in distributed workloads on Kubernetes.

The significance of these findings is fourfold:
\begin{enumerate}
    \item \textbf{Quantifiable Performance Impact}: We have quantified the substantial performance penalty of suboptimal hardware placement.
    \item \textbf{Architectural Simplification}: The KND model provides a robust, simplified, and operationally superior mechanism for achieving this performance.
    \item \textbf{A Foundation for Broader High-Performance Applications}: The principles demonstrated here extend far beyond GPU workloads. The ability to expose qualitative attributes like NUMA node is critical for other latency-sensitive domains, most notably telecommunications. For Network Functions Virtualization (NFV) workloads, ensuring that a VNF's vCPUs, its memory, and the physical NIC it uses all reside on the same NUMA node is paramount. CPU pinning, combined with NUMA alignment, is a known strategy for reducing jitter and achieving the deterministic, low-latency performance that these applications require.
    \item \textbf{Operational and Hardware Efficiency}: The pod startup latencies in the order of a few seconds enabled by this model offer a profound efficiency gain when contrasted with traditional infrastructure management. Provisioning new virtual machines to satisfy specific hardware topologies is a slow process, often on the order of minutes. This approach drastically improves hardware utilization and reuse, as the same underlying nodes can serve different workload profiles without requiring administrators to perform slow, manual infrastructure changes.
\end{enumerate}

\section{Future Work}

The KND model provides a flexible and extensible foundation for a new generation of networking capabilities in Kubernetes. Rather than extending the model itself, our future work will focus on fostering the ecosystem of drivers that this new paradigm enables, proceeding along several parallel tracks:

\begin{enumerate}
    \item \textbf{Scalability and Hardware Efficiency}: We will first address the limitations of the current study by evaluating the performance of topologically aligned resources on larger clusters. Beyond simple scaling, we will investigate the hardware efficiency improvements enabled by scale-out deployment patterns. This research will explore how the KND model facilitates breaking down monolithic, resource-intensive jobs into smaller, concurrent pods. By enabling fine-grained, topology-aware allocation of resources (e.g., assigning four pods with two GPUs each versus one pod with eight GPUs on a single node), we hypothesize that overall cluster throughput and hardware utilization can be significantly increased, providing a more cost-effective approach for large-scale AI and HPC environments.

    \item \textbf{An Interoperable Ecosystem for Telco and Advanced Networking}: The KND model is uniquely positioned to address the complex demands of telecommunications workloads. Our future work will focus on enabling a "galaxy" of independent, composable drivers that cater to this domain on two distinct fronts:

  \begin{itemize}
        \item \textbf{Data Plane Optimization}: Building on the model's ability to expose NUMA data, we will enhance DraNet to support explicit CPU pinning strategies. A future driver could allow a ResourceClaim to request not only a NIC on a specific NUMA node but also to pin the pod's workload threads to the CPU cores on that same node, providing a fully aligned, high-performance data path.
        \item \textbf{Control Plane and Service Integration}: The model's true power lies in managing abstract services. We envision specialized drivers emerging that manage complex carrier-grade network protocols. For instance, a driver could advertise connectivity to MPLS or SRv6 VPNs, or even to specific slices of a 5G Radio Access Network (RAN). A user could then claim "access to the 5G core network slice," and that driver would perform the necessary BGP peering, route leaking, or protocol-specific setup to connect the application.
    \end{itemize}
    Crucially, this ecosystem of independent drivers can be coordinated through standardized APIs. The ongoing work in KEP-4817 \cite{KEP4817} to standardize the network data reported in the ResourceClaim status is the key enabler for this interoperability. By ensuring all drivers report allocated interface details in a common format, it becomes possible for different network drivers to be composed together, fulfilling the ultimate vision of a truly network-aware, cloud-native ecosystem.

\end{enumerate}

\section{Conclusion}
The traditional methods for managing high-performance networking in Kubernetes are fraught with operational complexity and lack the expressiveness needed for modern, topology-sensitive hardware. The Kubernetes Network Driver (KND) model, built on the foundational APIs of DRA and NRI, offers a path forward. Our reference implementation, DraNet, demonstrates the tangible benefits of this model. It simplifies the deployment of RDMA networking and, most importantly, enables performance-aware scheduling by exposing crucial hardware topology attributes to Kubernetes. Our experiments quantitatively prove that this architectural shift translates directly into significant performance gains for real-world distributed workloads. By embracing a truly cloud-native approach to resource management, the KND model paves the way for the next generation of high-performance computing on Kubernetes.

\section*{Acknowledgments}
This research greatly benefited from the collaborative spirit and invaluable contributions of various Open Source communities. I extend my sincere gratitude to the Kubernetes SIG Network community, especially my co-leads Tim Hockin from Google and Dan Winship from Red Hat, for their guidance, insights, and relentless dedication to advancing Kubernetes networking. My thanks also go to the Kubernetes Device Management Working Group for their foundational work in resource management, and to Lionel Jouin from Ericsson Software Technology for his insightful collaboration and contributions at the intersection of DRA and networking. The ongoing efforts of the CNI community, the Open Container Initiative (OCI), and the communities behind controller runtimes are equally appreciated, as their foundational work on runtime specifications is crucial for the advancements presented in this paper. This work would not have been possible without the collective knowledge and supportive environment of these vibrant Open Source ecosystems.

\end{document}